\documentclass[reprint,twocolumn,aps,prb,amsmath,amssymb,floatfix,superscriptaddress,longbibliography]{revtex4-2}
\usepackage{graphicx}
\usepackage{epsfig}
\usepackage{bm}
\usepackage{dcolumn}
\usepackage{color}
\usepackage{physics}
\usepackage{float}
\usepackage[colorlinks,urlcolor=blue,citecolor=blue,linkcolor=magenta]{hyperref}
\makeatletter
\newcommand*{\rom}[1]{\expandafter\@slowromancap\romannumeral #1@}
\makeatother
\usepackage{tikz}
\usepackage{subfigure}

\begin{document}
	
\preprint{APS/123-QED}
\preprint{This line only printed with preprint option}
	
\title{Emergent dynamical quantum phase transition in a $Z_3$ symmetric chiral clock model}

\author{Ling-Feng Yu}
\affiliation{Key Laboratory of Atomic and Subatomic Structure and Quantum Control (Ministry of Education), Guangdong Basic Research Center of Excellence for Structure and Fundamental Interactions of Matter, South China Normal University, Guangzhou 510006, China}
\affiliation{Guangdong Provincial Key Laboratory of Quantum Engineering and Quantum Materials,School of Physics, South China Normal University, Guangzhou 510006, China}

\author{Wei-Lin Li}
\affiliation{Key Laboratory of Atomic and Subatomic Structure and Quantum Control (Ministry of Education), Guangdong Basic Research Center of Excellence for Structure and Fundamental Interactions of Matter, South China Normal University, Guangzhou 510006, China}
\affiliation{Guangdong Provincial Key Laboratory of Quantum Engineering and Quantum Materials,School of Physics, South China Normal University, Guangzhou 510006, China}

\author{Xue-Jia Yu}
\email{xuejiayu@fzu.edu.cn}
\affiliation{Department of Physics, Fuzhou University, Fuzhou 350116, Fujian, China}
\affiliation{Fujian Key Laboratory of Quantum Information and Quantum Optics,College of Physics and Information Engineering,Fuzhou University, Fuzhou, Fujian 350108, China}

\author{Zhi Li}
\email{lizphys@m.scnu.edu.cn}
\affiliation{Key Laboratory of Atomic and Subatomic Structure and Quantum Control (Ministry of Education), Guangdong Basic Research Center of Excellence for Structure and Fundamental Interactions of Matter, South China Normal University, Guangzhou 510006, China}
\affiliation{Guangdong Provincial Key Laboratory of Quantum Engineering and Quantum Materials,School of Physics, South China Normal University, Guangzhou 510006, China}

\date{\today}

\begin{abstract}
We study the quench dynamics in a $Z_3$ symmetric chiral clock model (CCM). The results reveal that chiral phases can lead to the emergence of dynamical quantum phase transition (DQPT). By analyzing Lee-Yang-Fisher zeros' distribution in the complex plane, we uncover the relation between the chiral phase and the emergence of DQPT. In concrete terms, only by taking some special angles can DQPT be induced. We confirm the above relation by computing the non-analytic points in Loschmidt echo return rate function. Furthermore, through the analysis of the corresponding dynamical partition function, we reveal the mechanism of the emergent DQPT and deduce the analytical expression of dynamical partition function's zero points' coordinates. Based on the analytic expression, one can obtain all the angles that induce DQPT's emergence and predict more possible DQPT in the system.
\end{abstract}

\maketitle

\section{Introduction}
In 1952, Chen-Ning Yang and Tsung-Dao Lee published two pioneering works which respectively put forward the concept of Yang-Li theory~\cite{CNYang1952} and Li-Yang zero~\cite{TDLee1952}, providing a solid basis for understanding the equilibrium phase transition. In recent years, with the deepening of the study of phase transitions, non-equilibrium phase transitions have stepped into the spotlight~\cite{TAo2006,GDagvadorj2015,FTurci2017,CKlockner2020}. For instance, dynamical quantum phase transition (DQPT), as a dynamical extension of thermodynamic phase transition, has received extensive attention. DQPT usually occurs during quench dynamics across phase transition points~\cite{MHeyl2013,MHeyl2018_review,MHeyl2019}. Unlike the free energy in thermodynamic phase transitions, the related dynamical free energy in dynamical quantum transitions is dependent on imaginary time rather than temperature. The corresponding dynamical partition function shows the inner product of the final state and the initial state in the quench dynamics, which is also known as the return amplitude. Under the thermodynamic limit, when the return amplitude is distributed in the imaginary axis, the corresponding dynamical free energy will have non-analytic points, and then DQPT will emerge in the system~\cite{MHeyl2013}.

Since M. Heyl first proposed DQPT in 2013, a series of milestone results have been achieved in DQPT-related fields over the past decade~\cite{HTQuan2006,MHeyl2013,JCBudich2016,MHeyl2017_speedLimits,Bhattacharya2017,IHagymasi2019,XNie2020,Mishra_2020,RJafari2021,MSadrzadeh2021,JNaji2022,YZeng2023,ONKuliashov2023}. At first, it was thought that this phenomenon would only occur in the quench dynamics across the traditional critical point~\cite{MHeyl2013,CKarrasch2013,MHeyl2015,CKarrasch2017,MHeyl2018}. However, with the deepening of the research, people have recently discovered abnormal DQPT phenomena, such as DQPT that crosses the topological critical point~\cite{MHeyl2017,VVijayan2023} and some DQPTs that do not need to cross the critical point~\cite{TYZou2023, YHHuang2024}. At present, DQPT has been experimentally observed in a variety of artificial quantum simulators. Examples include optical lattice ultracold atomic systems~\cite{Flaschner2018}, trapped ions~\cite{PJurcevic2017}, superconducting circuits~\cite{GXueYi2019}, Rydberg atomic arrays~\cite{Bernien2017}, etc~\cite{Sharma2016,Zhang2017,WKunkun2019,TTian2019,Xu2020}. Moreover, the characterization of DQPT from different angles has also been recently discussed, for example, the characterization of DQPT by dynamical order parameters~\cite{JMeibohm2022}, and the characterization of DQPT by the out-of-time-ordered correlator~\cite{XFNie2020,WLZhao2021,WLZhao2022,WLZhao2023}. With the deepening of the research in the non-Hermitian field~\cite{SYYao2017,SYYao2018,SYYao2018_2,FSong2019,FSong2019_2,DWZhang2020,EJBergholtz2021,AMGarcia2022,YSun2023,KKawabata2023,SZLi2024,SZLi2024_2,SZLi2024_3,GJLiu2024}, more and more attention has recently been paid to DQPT in non-Hermitian systems~\cite{LWZhou2018,DMondal2022,DMondal2023,YCJing2024}.

On the other hand, chiral clock model (CCM), as one of the most typical platforms for demonstrating many-body quantum phase transitions and critical behaviors, has attracted extensive attention in recent years~\cite{DAHuse1981,Ostlund1981,DAHuse1982,Haldane1983,HUSE1983,HOWES1983,YZhuang2015,RSamajdar2018,NNishad2021}. In its nature, CCM can be seen as introducing an additional chiral phase into the $N$-state Potts model. Specifically, the $Z_3$ chiral clock model we focus on in this paper can actually be regarded as introducing a chiral phase into the $3$-state Potts model~\cite{SOstlund1981,DAHuse1981,FPaul2012}. The introduced chiral phase will lead to the occurrence of the commensurate-incommensurate phase transition~\cite{Ostlund1981,HOWES1983}, which will bring about interesting domain wall dynamics and entanglement properties in the system~\cite{YZhuang2015,NNishad2021,XJYU2023}.

So far, although a lot of research has been done on CCM or Potts models, few work has focused on DQPT in such systems. In this paper, we will devote to emploring DQPT in $Z_3$ symmetric CCM. Notably, by the authors of reference~\cite{CKarrasch2017} demonstrated in 2017 that DQPT does not occur during the quench dynamics of the $Z_3$ Potts model from a fully polarized ferromagnetic (FM) phase to a paramagnetic (PM) phase~\cite{CKarrasch2017}. Here, we will show that the additional chiral phase will cause DQPT to emerge in CCM.

The rest of this paper is organized as follows. In Sec.~\ref{S2}, we briefly introduce the chiral clock model of $Z_3$ symmetry, and then analytically calculate the quench dynamics from the FM phase to the PM phase. We illustrate with examples that DQPT can only be induced by special chiral phases. In Sec.~\ref{S3}, we discuss the corresponding Fisher zeros with different chiral phases, and further demonstrate the above conclusion. In Sec.~\ref{S4}, we discuss the reason why these particular phases can trigger DQPT by decomposing the dynamical partition function. The main findings of this paper are summarized in Sec.~\ref{S5}.

\section{model}\label{S2} 
\begin{figure}[htbp] \centering
	\includegraphics[width=8.3cm]{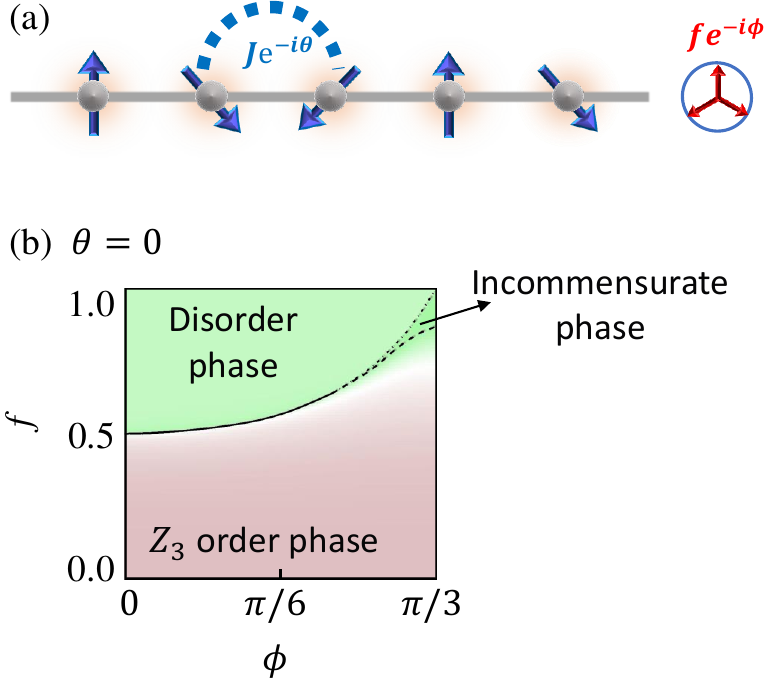}\\
	\caption{(a) The schematic representation of  $Z_3$ symmetric chiral clock model. (b) The corresponding phase diagram.}
	\label{F1}
\end{figure}

We start at a $Z_3$ symmetric chiral clock model, which is essentially equivalent to introducing a phase modulated term into a 3-state Potts model [see Fig.~\ref{F1}(a)]. The corresponding Hamiltonian reads,
%%%
%%% 
\begin{equation}\label{Hami}
H=-J\sum_{j=1}^{N-1}\sigma_j^\dagger \sigma_{j+1}e^{-i\theta}-f\sum_{j=1}^{N}\tau_j^\dagger e^{-i\phi}+H.c.,
\end{equation}
%%%
%%%
where $J$ and $f$ denote the coupling coefficient and the transverse field strength, respectively. Without loss of generality, we set $J=1-f$, where $f\in [0,1]$. Then, one can find that the system behaves as a FM (PM) phase under the condition of $f=0$ ($f=1$). $\sigma_j$ and $\tau_j$ denote local spin operators on site $j$, which satisfy $Z_3$ symmetry, i.e., $\sigma_j^3=\tau_j^3=1$ and $\sigma_j \tau_j = \omega \tau_j \sigma_j$, where $\omega=e^{i2\pi/3}$. They are represented by 3 dimension matrices, i.e.,
%%%%%%
%%%%%%
%%%%%%
%%%%%%

%%%
%%%
\begin{equation}\label{CCM_Operator}
	\sigma_j =  \begin{pmatrix}
	 	0 & 1 & 0  \\
	 	0 & 0 & 1  \\
	 	1 & 0 & 0  \\
	\end{pmatrix}, 
 	\tau_j =  \begin{pmatrix}
	1 & 0      & 0        \\
	0 & \omega & 0        \\
	0 & 0      & \omega^2 \\
	\end{pmatrix}.
\end{equation}
%%%
%%%
In the $\tau$ representation, $\sigma_j^\dagger \sigma_{j+1}$ describes the spin-spin interaction between sites $j$ and $j+1$, whereas $\tau_j$ denotes the local on-site potential on $j$th site. The spin operator on each $j$ site has a three-dimensional subspace, and one can label them as $|1\rangle$, $|\omega\rangle$, $|\omega^2\rangle$. $\theta$ and $\phi$ are two different chiral phases as shown in Fig.~\ref{F1}(a). The model exhibits three distinct phases, including a $Z_3$ ordered phase ferromagnetic phase, an incommensurate floating phase and a disordered paramagnetic phase. By computing the entanglement entropy or the central charges versus $f$ and $\phi$, one can obtain the corresponding phase diagram [see Fig.~\ref{F1}(b)].

\begin{figure}[htbp] \centering
	\includegraphics[width=8.3cm]{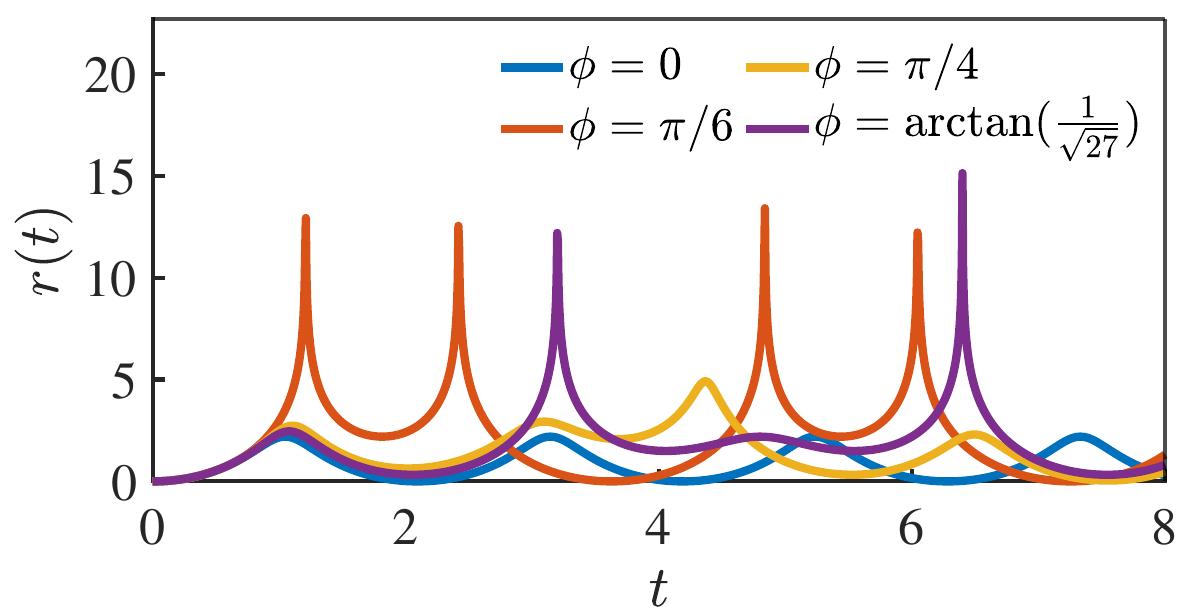}\\
	\caption{The rate functions over time under the quench conditions from a FM phase to a PM phase. Throughtout, $\theta=0$.}
	\label{fig2}
\end{figure}

The Loschmidt echo return rate function (rate function for short), as the key quantity characterizing DQPTs, can be defined as
%%%
%%% 
\begin{equation}\label{RF}
r(t)=-\frac{1}{N}\ln{|G(t)|^2},
\end{equation}
%%%
%%% 
where 
%%%
%%% 
\begin{equation}\label{RA1} 
G(t)=\bra{\psi_0}e^{-i Ht}\ket{\psi_0}
\end{equation}
%%%
%%% 
is the dynamical partition function (also called return amplitude), which exhibits the probability amplitude of the inner product between the final state and the initial state~\cite{MHeyl2019}. Previous studies have shown that if there are rate function's non-analytic points during a quench dynamical process, it indicates the emergence of DQPT~\cite{MHeyl2013}. Without loss of generality, we consider a typical case ($\theta=0$) for simplicity, i.e., a quench process from a FM phase ($f=0$) to a PM phase ($f=1$). A fully polarized pure state is considered as the initial state, which can be represented as
%%%
%%% 
\begin{equation}\label{Int}
	|\psi_0\rangle=(3^{-\frac{N}{2}})\prod_{j=1}^{N}\left(\sum_{m=1}^{3}{|\omega^{m-1}\rangle_j}\right).
\end{equation}
%%%
%%% 
The corresponding Hamiltonian of PM phase reads
%%%
\begin{equation}\label{E6}
H_{\text{PM}} = -f\sum_{j=1}^{N}\tau_{j}^{\dagger} e^{-i\phi} +H.c.
\end{equation}
%%%
%%%
From the expressions~\eqref{Int} and~\eqref{E6}, one can get two important pieces of information. First, the initial state of the system is composed of direct product states of lattice points. Second, since there is no coupling between sites in the Hamiltonian corresponding to PM phase ($f=1$), the quench dynamics is limited to isolated sites. In other words, the initial state and time evolution are exactly the same for different system size $N$. Besides, the dynamic partition function is used to describe dynamic properties of the system in non-equilibrium state. Since the dynamic behavior of different system sizes is the same, $N$ has no effect on the dynamic partition function, and likewise, $N$ has no effect on the rate function. Therefore, we can safely set $N=1$ in the following calculation. The result obtained based on this is consistent with the result of taking an arbitrary value of $N$ (even $N\rightarrow\infty$). Then, the corresponding rate function can be obtained as
%%%
%%% 
\begin{equation}\label{E7}
    r(t)=-\ln{\left|\frac{1}{3}\sum_{m=1}^{3} \exp[-it(\omega^{(m-1)}e^{i\phi}+\omega^{-(m-1)}e^{-i\phi})]\right|^2}.
\end{equation}
%%%
%%% 
We plot Eq.~\eqref{E7} in Fig.~\ref{fig2}. For the case of $\phi=0$, i.e., without chiral phases, the model degenerates into the $Z_3$ Potts model. It is evident that the rate function is smooth across each temporal period, with no occurrence of DQPT.
Generally speaking, the introduction of a phase, such as $\phi=\frac{\pi}{4}$, maintains the smoothness of the rate function and does not trigger any DQPT.
Only when special phases are introduced, such as $ \phi = \frac{\pi}{6} $ or $ \phi = \arctan\left(\frac{1}{\sqrt{27}}\right) $, does the rate function exhibit non-analytical behavior (kinks) in its time evolution, indicating non-analytic points that trigger DQPTs [see Fig.~\ref{fig2}].
%%%%%%
%%%%%%
%%%%%%
%%%%%%

\section{Fisher zeros}\label{S3}
In 1965, Michael E. Fisher extended the Yang-Lee theory to finite temperature systems~\cite{MEFisher1965}. In other words, he extended the Lee-Yang zero's concept to the temperature’s complex plane. In honor of this pioneering work, the corresponding zero points were named Fisher zeros. In 2013, Markus Heyl noticed that the dynamical evolution operator of a quantum system has mathematical similarity to the partition function of a finite temperature system, i.e., by taking the imaginary time $it$ as the inverse of temperature $\beta$ in the partition function, one can use Fisher's theory to analysis DQPT~\cite{MEFisher1965}. In concrete terms, the quantity here we pay attention to is boundary partition function in the complex plane~\cite{ALeClair1995}, i.e.,
%%%
%%% 
\begin{equation} \label{FPF}
	z(\beta) = \bra{\psi_j}e^{-\beta H}\ket{\psi_j},
\end{equation}
%%%
%%%
where $\beta\in\mathbb{C}$. The partition function here reveals different physics by the real and imaginary values of $\beta$. On the one hand, for real $\beta=R$, Eq.~\eqref{FPF} can be interpreted as the partition function of the field theory described by $H$ with boundaries described by boundary states $\psi_j$ separated by $R$. On the other hand, for imaginary $\beta=it$, the corresponding expression of $z(\beta)=G(t)$, which only gives Eq.~\eqref{RA1}~\cite{MHeyl2018_review}.

According to Fisher's theory, the falling of Fisher zeros on the real and imaginary axes corresponds to the occurrence of equilibrium phase transition and DQPT, respectively. We analytically derive the expression of dynamical partition function, i.e.,
%%%
%%% 
\begin{equation}\label{E9}
	z(\beta)=\frac{1}{3}\sum_{m=1}^{3} \exp[-\beta(\omega^{(m-1)}e^{i\phi}+\omega^{-(m-1)}e^{-i\phi})].
\end{equation}
%%%
%%% 

%%%%%%
%%%%%%
\begin{figure}[htbp] \centering
	\includegraphics[width=8.3cm]{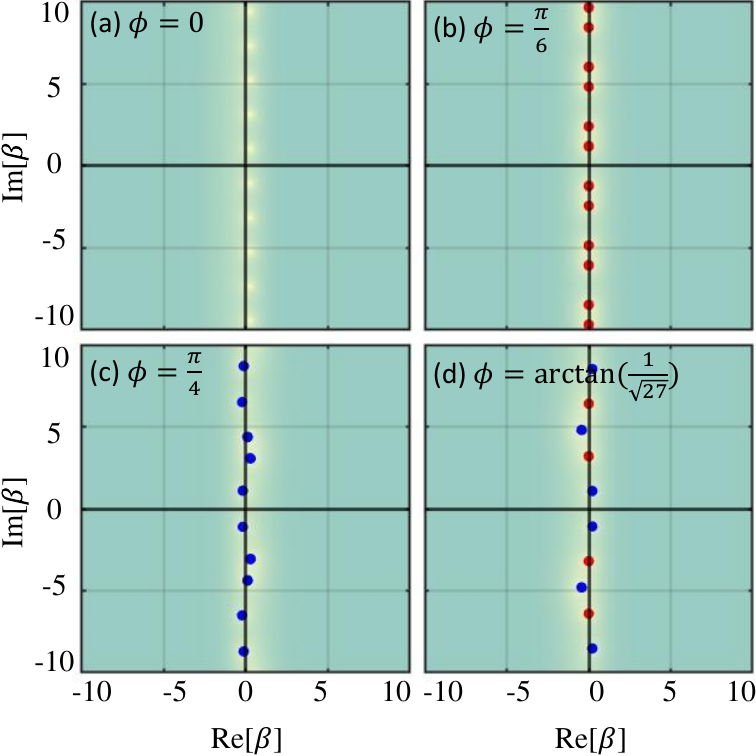}\\
	\caption{The distribution of Fisher zeros in the complex plane with chiral phases $\phi=0,\frac{\pi}{6},\frac{\pi}{4},\arctan{\frac{1}{\sqrt{27}}}$. The red and blue dots represent Fisher zeros that fall and do not fall on the imaginary axis, respectively.}
	\label{F3}
\end{figure}
%%%%%%
%%%%%%
Through the expression~\eqref{E9}, we obtain the distribution of Fisher zeros in the complex plane (see Fig.~\ref{F3}). As shown in the figure, when $\phi=0$, the system will be reduced to the standard 3-state Potts model. The corresponding results exhibit that no Fisher zero appears [see Fig.~\ref{F3}(a)]. That is to say, DQPT does not occur in the quench dynamics, which is consistent with the results proved by the authors of reference~\cite{CKarrasch2017} in 2017. Furthermore, when some special angles are taken for $\phi$, for example, $\phi=\pi/6$ or $\phi=\arctan(1/\sqrt{27})$, Fisher zeros will appear and fall on the imaginary axis (the red dots). Under these conditions, DQPT will emerge in the system [see Fig.~\ref{F3}(b)(d)]. While one can see the emergence of DQPT at certain special angles, most $\phi$ cannot give rise to Fisher zero on the imaginary axis (the blue dots). In other words, in most cases, DQPT cannot be induced even if the chiral phase is introduced. Naturally we are curious that, how can we derive those phases capable of inducing DQPT?

\section{Dynamical partition functions}\label{S4}
\begin{figure}[htbp] \centering
	\includegraphics[width=8.3cm]{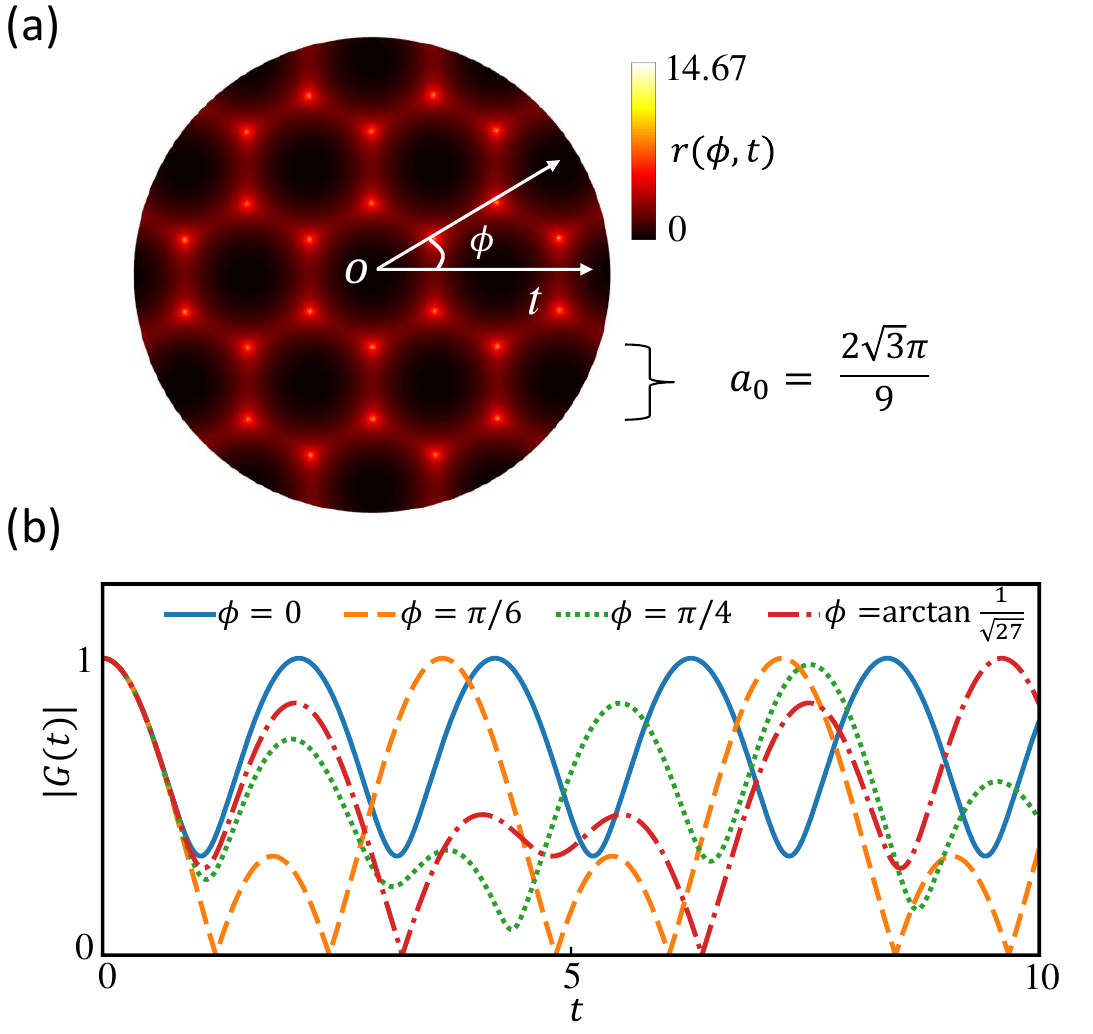}\\
	\caption{Panoramic rate function (a) and the corresponding non-analytic points versus $\phi$ and $t$. Panel (b) presents the curves of the return amplitude $G$ versus $t$.}
	\label{F4}
\end{figure}

To answer the above question, we plot the panoramic diagram of rate function Eq.~\eqref{RF} with respect to chiral phases $\phi$ and time $t$ in the polar coordinates~[Fig.~\ref{F4}(a)]. The results exhibit that the distribution of the non-analytic points form a ``honeycomb lattice'' structure with the ``lattice constant'' of $a_0=\frac{2\sqrt{3}\pi}{9}$. Since the non-analytic points of the rate function all fall on the vertex of the ``hexagonal lattice'' structure, the emergence of DQPT in the quench dynamics is possible only when a suitable phase angle is selected.

%%%%%%
%%%%%%
%%%%%%
%%%%%%

To further reveal the conditions for the emerged DQPT, we analyze the return amplitude $G(t)$ below. From Eq.~\eqref{E6}, one can get the Hamiltonian of the PM phase ($f=1$), i.e.,
%%%
%%%
\begin{equation}\label{H_pm}
	H_{\text{PM}} =  \begin{pmatrix}
		2\cos\left(\phi\right) & 0 & 0  \\
		0 &      2\cos\left(\phi + \frac{2\pi}{3}\right) & 0  \\
		0 & 0 &  2\cos\left(\phi + \frac{4\pi}{3}\right)  \\
	\end{pmatrix}.
\end{equation}
%%%
%%%
It is not difficult to find that the matrix is zero except for the elements on the diagonal. In other words, there is only on-site operator in the quench process. The fully polarized pure state of Eq.~\eqref{Int} is chosen as the initial state of quench dynamics. Under the condition, one can derive the analytical expression of the dynamical partition function with respect to $\phi$ and $t$, i.e.,
%%%
%%%
\begin{equation}\label{RA2}
  G(\phi,t)=\frac{e^{-i2t\cos{\phi}}}{3}\left[e^{i 2\sqrt{3} t \cos(\phi-\frac{\pi}{6})} + e^{i 2\sqrt{3} t \cos(\phi+\frac{\pi}{6})} + 1 \right].
\end{equation}
%%%
%%% 
We show the behaviors of dynamical partition functions through analytic expression~\eqref{RA2} for different $\phi$. 

First, let's consider the case of $\phi=0$. The corresponding dynamical partition function reads
%%%
%%% 
\begin{equation}\label{returnAmplitude_0_t}
	G(0,t)=\frac{e^{-i2t}}{3} \left(e^{i3t}+e^{i3t}+1 \right).
\end{equation}
%%%
%%% 
It can be seen clearly that $\left| G(0,t) \right| \geq \frac{1}{3}$, which means, no matter what time $t$ evolves to, the dynamical partition function can never be equal to zero, so DQPT can never happen in this case~[see blue solid line in Fig.~\ref{F4}(b)]. This once again proves that the conclusion of reference~\cite{CKarrasch2017} is correct~\cite{CKarrasch2017}.

Next, let's have a look at what will happen after the chiral phases have been introduced. Under the condition of $\phi=\frac{\pi}{6},\frac{\pi}{4},\arctan(\frac{1}{\sqrt{27}})$, one can obtain the corresponding expressions for dynamical partition function, respectively, i.e.,
%%%
%%% 
\begin{equation}\label{RAP6}
	G(\frac{\pi}{6},t) = \frac{e^{-it\sqrt{3}}}{3} \left[e^{it 2\sqrt{3}}+e^{it \sqrt{3}}  + 1 \right],
\end{equation}
%%%
%%% 
%%%
%%% 
\begin{equation}\label{RAP4}
	G(\frac{\pi}{4},t) = \frac{e^{-it\sqrt{2}}}{3} \left[e^{i t \frac{\sqrt{6}+3\sqrt{2}}{2}}+e^{i t \frac{\sqrt{6}-3\sqrt{2}}{2}}  + 1 \right],
\end{equation}
%%%
%%% 
%%%
%%% 
\begin{equation}\label{RA27}
	G(\arctan{\frac{1}{\sqrt{27}}},t) = \frac{e^{-it\frac{3\sqrt{21}}{7}}}{3}
	  \left[
		e^{ i t  \frac{5\sqrt{21}}{7}}
		+e^{ i t  \frac{4\sqrt{21}}{7}}
		+ 1
		\right].
\end{equation}
%%%
%%% 
We plot the results of ~\eqref{RAP6} (orange dashed line),~\eqref{RAP4} (green dotted line) and~\eqref{RA27} (red dash-dotted line) in Fig.~\ref{F4}(b), respectively. The results show that under certain circumstances (such as $\phi=\pi/6$ or $\arctan{\frac{1}{\sqrt{27}}}$), the $G(t)$ curve versus time will touch zero. Then, the corresponding rate function will have points with non-analytical behavior, so DQPT emerges. On the other hand, in most cases (such as $\phi=\pi/4$), the dynamical partition function can never fall on zero, as a result, the DQPT will not occur.

Let's further analyze why DQPT occurs by splitting up the $G(\phi,t)$. One can divide the dynamical partition function $G(\phi,t)$ into three parts, i.e.,
%%%
%%% 
\begin{equation}\label{returnAmplitude_expand_t}
G\left(\phi,t\right) = \sum_{m=1}^{3}G_{m}(\phi,t) =\sum_{m=1}^{3}\frac{e^{-i t E_m}}{3}  \langle  \psi_m \mid  \psi_0 \rangle.
\end{equation}
%%%
%%% 
$|\psi_{m=1,2,3}\rangle=|1\rangle$,~$|\omega\rangle$,~$|\omega^2\rangle$ are the basis vectors in the spinor subspace and the corresponding energy $E_m = 2\cos\left[\phi + (m-1)\frac{2\pi}{3}\right]$ as depicted in Eq.~\eqref{H_pm}.

%%%%%%
%%%%%%
\begin{figure}[htbp] \centering
	\includegraphics[width=8.3cm]{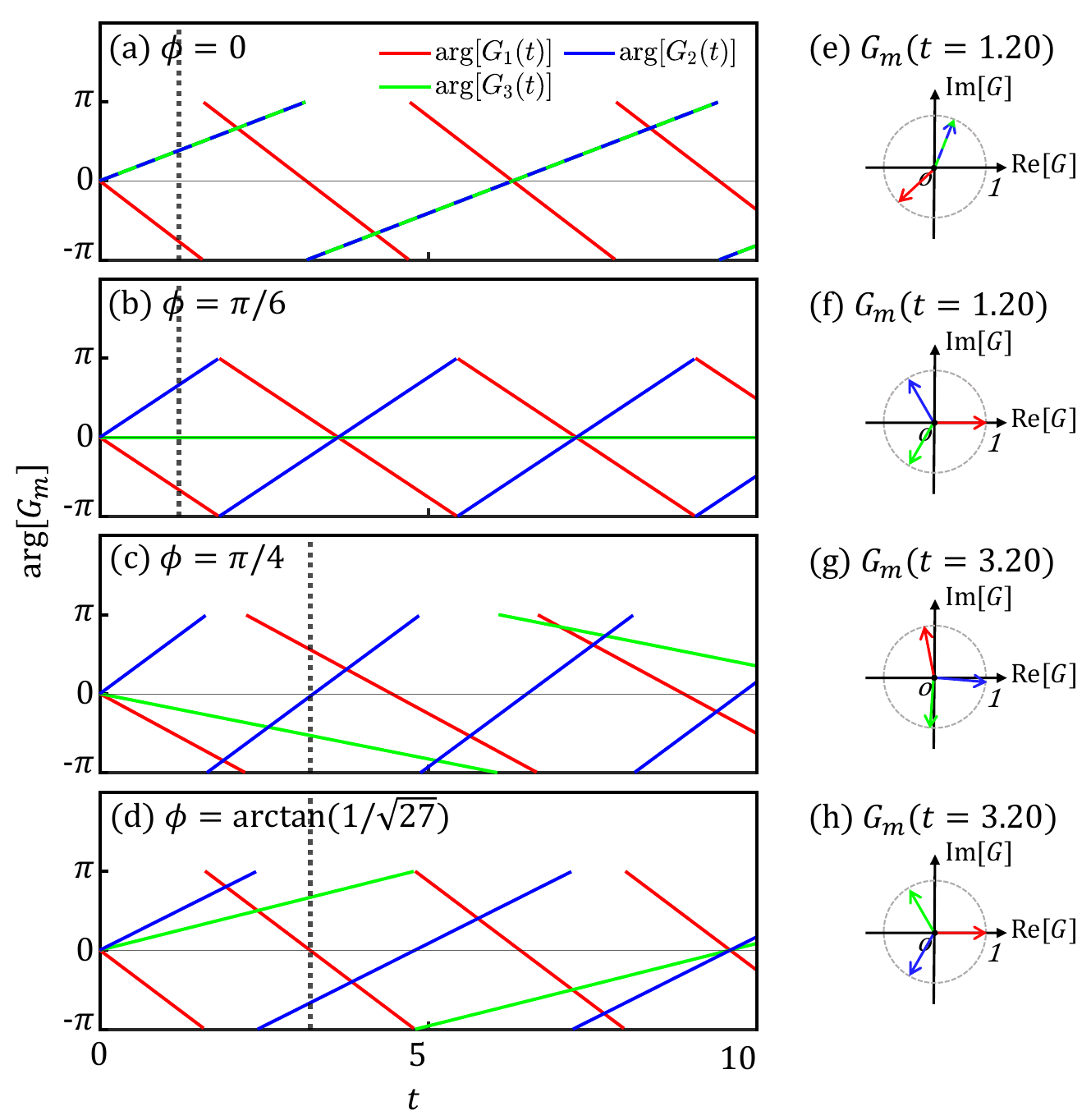}\\
	\caption{Subcomponents' arguments $\arg[G_m]$ versus $t$ with $\phi=0$ (a),~$\pi/6$~(b),~$\pi/4$~(c),~$\arctan{\frac{1}{\sqrt{27}}}$~(d). (e-h) Visualized summation $\sum_{m=1}^{3}G_m$ in complex plane.}
	\label{F5}
\end{figure}
%%%%%%
%%%%%%

Note that, the modulus of $G_{m=1,2,3}$ are the same, all equal to $1/\sqrt{3}$. $G_{m=1,2,3}$ is a periodic function with period of $2\pi/E_m$. The arguments of subspace dynamical partition function $\arg{[G_m]}$ show periodic evolution over time~[see Fig.\ref{F5}(a)-(d)]. Then, by changing the chiral phase $\phi$, $E_m$ can be adjusted to affect the behavior of the $G_{m}(t)$ curve. For example, in the case of introducing different phase angles $\phi$, the values of $G_{m=1,2,3}$ will rotate in the complex plane~[see Fig.\ref{F5}(e)-(h)]. Since the modulus are equal, when the three phase angles form a certain relationship, the vectors' sum of the three complex numbers will be zero, i.e., they satisfy the equation
%%%
%%% 
\begin{equation}\label{zeros_equation}
	\sum_{m=1}^{3} e^{-2i t \cos\left[\phi+(m-1)\frac{2\pi}{3}\right]}=0.
\end{equation}
%%%
%%%
In other words, after fixing some specific chiral phase angle $\phi$, the time t satisfying the above expression will lead to the emergence of DQPT. These specific times are often called the critical time, denoted as $t^*$. For example, here we provide four typical chiral phase angles and calculate the corresponding $\arg[G_{m}(t)]$. The result reveals that for the case of $\phi=0$ and $\phi=\pi/4$, the arguments cannot form a suitable composite pattern at any time to make the sum of $G_1,G_2,G_3$ be zero in the complex plane. However, when $\phi=\pi/6$ or $\phi=\arctan{1/\sqrt{27}}$, the summation of $G_1,G_2,G_3$ can be zero at $t^*$, resulting in the dynamical partition function $G(\phi,t)=0$. Therefore, in these cases, the system will have DQPT. The analytical results here agree well with the results of Fisher Zero analysis, which once again prove that DQPT can be induced by introducing the chiral phase.

Now, we discuss the rules for the selection of chiral phases. Mathematically, by analyzing the equation $G(\phi,t)=0$, we obtain that if DQPT emerges, $\phi$ and $t$ must satisfy the following two analytic expressions, i.e.,
%%%
%%%
\begin{equation}\label{phipq}
\phi(p,q)=2\arctan{\frac{2\sqrt{p^2+q^2+pq}-\sqrt{3}p}{p+2q}},
\end{equation}
%%%
%%%
and
%%%
%%%
\begin{equation}\label{tpq}
t(p,q) = \frac{2\pi\sqrt{p^2+q^2+pq}}{3\sqrt{3}},
\end{equation}
%%%
%%%
where $(p,q)\in \mathbb{Z}$ and $p-q\mod3\neq0$. 

From the above expressions~\eqref{phipq} and \eqref{tpq}, one can obtain the position where zeros appear in the two dimensional polar coordinate plane. In other words, the $\phi$ and $t$ corresponding to any integer pair $(p,q)$ are the coordinates of the nonanalytic points of DQPT. However, for some phases $\phi$ that have no integer pair $(p,q)$ to correspond, DQPT can never be induced (e.g. $\phi=0$ or $\phi=\pi/4$).

Specifically, when $(p,q)=(1,0)$, one can obtain $\phi(1,0)=\pi/6$, which makes the expression~\eqref{phipq} valid. Besides, $\phi(3,-1)=\arctan{1/\sqrt{27}}$ can also satisfy the above analytical expression~\eqref{phipq}. For $\phi=\pi/4$ or $\phi=0$, no suitable integer pair $(p,q)$ can be found to meet the condition. That is to say, zero point of the dynamical partition function appears only when $\phi$ satisfies the expression~\eqref{phipq}, such as $\phi=\pi/6$ or $\arctan{1/\sqrt{27}}$. 

Note that, the expression~\eqref{phipq} and~\eqref{tpq} are universal, therefore, one can find more chiral phases that can induce DQPT by providing different sets of $(p,q)$.

\section{Summary}\label{S5} 
In conclusion, the quench dynamics from FM to PM is studied in the $Z_3$ symmetric CCM. The results show that DQPT will emerge when specific chiral phases angle is introduced. Furthermore, we prove our conclusion through the behavior of Fisher zeros and dynamical partition function. Based on the idea of dividing dynamical partition function, we obtain the analytic expressions of the phase and time corresponding to the critical point position of DQPT. Future intriguing questions include analyzing the distribution of Lee-Yang-Fisher zeros and the dynamical partition function in higher dimensions and within different symmetry groups (e.g., $Z_4$, U(1), among others), as well as examining the effects of finite temperature. Our work could offer new insights and provide a unique perspective for understanding DQPTs.

\begin{acknowledgments}
This work was supported by the National Key Research and Development Program of China (Grant No.2022YFA1405300) and the National Natural Science Foundation of China (Grant No.~12405034).
\end{acknowledgments}

\bibliography{main}
\end{document}